# The effects of point defect type, location, and density on the Schottky barrier height of Au/MoS$_2$ heterojunction: A first-principles study

Viacheslav Sorkin[1,*], Hangbo Zhou[1], Zhi Gen Yu[1], Kah-Wee Ang[2,3,†], Yong-Wei Zhang[1,‡]

[1]Institute of High-Performance Computing, A*STAR, 1 Fusionopolis Way, Singapore 138632

[2]Department of Electrical and Computer Engineering, National University of Singapore, 4 Engineering Drive 3, Singapore, 117583

[3]Institute of Materials Research and Engineering, A*STAR, 2 Fusionopolis Way, Singapore, 138634

## Abstract

Using DFT calculations, we investigate the effects of the type, location, and density of point defects in monolayer MoS$_2$ on electronic structures and Schottky barrier heights (SBH) of Au/MoS$_2$ heterojunction. Three types of point defects in monolayer MoS$_2$, that is, S monovacancy, S divacancy and Mo$_S$ (Mo substitution at S site) antisite defects, are considered. The following findings are revealed: (1) The SBH for the monolayer MoS$_2$ with defects is universally higher than that for its defect-free counterpart. (2) S divacancy and Mo$_S$ antisite defects increase the SBH to a larger extent than S monovacancy. (3) A defect located in the inner sublayer of MoS$_2$, which is adjacent to Au substrate, increases the SBH to a larger extent than that in the outer sublayer of MoS$_2$. (4) An increase in defect density increases the SBH. These findings indicate a large variation of SBH with the defect type, location, and concentration. We also compare our results with previously experimentally measured SBH for Au/MoS$_2$ contact and postulate possible reasons for the large differences among existing experimental

---

[*] Email: sorkinv@ihpc.a-star.edu.sg
[†] Email: eleakw@nus.edu.sg
[‡] Email: zhangyw@ihpc.a-star.edu.sg

**measurements and between experimental measurements and theoretical predictions. The findings and insights revealed here may provide practice guidelines for modulation and optimization of SBH in Au/MoS$_2$ and similar heterojunctions via defect engineering.**

# 1. Introduction

Metal-semiconductor junctions have been widely used in modern electronic devices. In such a junction, a Schottky barrier, which is a potential energy barrier for electron or hole, can be formed. The Schottky barrier height (SBH) is essential in rectifying electrical current characteristics [1]. Recently, a new type of computing devices based on artificial synapses (e.g., memtransistors, resistive synaptic switches, memristors, etc.) that mimic the biological neural systems have attracted significant research interests [2]. Of particular interest is the exploration of semiconducting two-dimensional (2D) materials for such artificial synapses, and molybdenum disulfide (MoS$_2$) monolayer, which is a typic semiconducting 2D material, is often used, and its junction with a metallic electrode becomes a principal building block [3–6]. Since SBH plays an important role in modulating charge carrier transport, switching characteristics [7] and device performance [8–10], accurately setting and adjusting the SBH is of critical importance for the control of charge transport in MoS$_2$ and the design of memory switching in MoS$_2$-based devices.

Yet, accurate control of SBH is still a challenge in designing semiconductor-based high-performance nanoscale electronics. It is known that many factors can affect the SBH of metal/MoS$_2$ junctions, such as strong Fermi-level pinning (FLP) [8,11–13], electronic band alignment at the interface [14,15], interface dipole moment formation due to the charge redistribution at the contact [11,16,17], bond formation between MoS$_2$ and the underlying substrate [18], push-back effect [10], work function of metals [11], dielectric screening due to MoS$_2$ layer [15], quantum confinement (the out-of-plane interactions between MoS$_2$ monolayer and underlying metallic strongly modifies the boundary condition for quantum confinement on one side of the MoS$_2$) [19], interfacial stress and strain [20,21] and the presence of defects in MoS$_2$ layer and metallic substrate (e.g., point and line defects of various types at different concentration and spatial distribution) [22–26]. Due to the complexity, there is a large scattering in existing experimental measurements of SBH, and there is also a large discrepancy between these experimental measurements and existing theoretical predictions [13–16]. Reconciliation between these discrepancies so far has not been achieved.

Two common techniques are employed to prepare MoS$_2$ samples. A commonly used one is via CVD growth [27]. Another commonly used one is mechanical exfoliation [28,29]. Compared with mechanically exfoliated samples, the CVD growth process, often taking place at relatively high temperatures, induces various native defects, including point defects, grain boundaries and edges [30]. The equilibrium concentration of point defects is determined by their formation energies and growth conditions (temperature, pressure, and chemical potential). Hence, experimentally observed defect densities vary strongly from experiment to experiment [26]. The estimated sulfur vacancy density is in the range $n_v$~$10^8$–$10^{11}$ cm$^{-2}$ [31–37].

The experimentally measured values of SBH for metal/MoS$_2$ interface often fall in a broad range. For example, the SBH for Au/MoS$_2$ contact is between 0.06 eV and 0.92 eV [1,8,38,39]. It is possible that the MoS$_2$ samples used in experiments are rather defective and inhomogeneous, which could result in the scattering and deviations from the intrinsic value of defect-free MoS$_2$/metal contact [36,40,41].

Moreover, it is common to apply the electrode deposition to create the metal/MoS$_2$ junction, in which the deposited ''high energy'' metal atoms can damage the crystal lattice of MoS$_2$. This deposition can lead to a substantial chemical disorder, namely formation of numerous S and Mo vacancies, and metallic-like defects (metallic impurities) at the interface [3,15,42]. The chemical disorder can have a profound effect on both the SBH and FLP. In contrast, when atomically flat metal thin films are laminated onto MoS$_2$ monolayer (without direct chemical bonding) by using the damage-free electrode transfer technique [1], the observed interface is effectively free from chemical disorder and FLP.

It is noted that the effect of the point defects on the SBH was studied by using DFT calculations, mostly focusing on the effects of S monovacancies on the electronic characteristics and SBH of the metal/MoS$_2$ contact [5, 11, 38, 39, 43-47]. For example, Feng et al. [12] studied the Ti/MoS$_2$ contact with S- and Mo-vacancies in MoS$_2$. They found that S vacancies reduced the SBH value, while the SBH vanished when Mo-vacancies were present, due to the stronger chemical interactions with the underlying substrate induced by defects. Yang et al. [23] found that the charge type of MoS$_2$ layer in the Au/MoS$_2$ contact could be tuned by adjusting the concentration of S-vacancies. At relatively low concentrations of S-vacancies, the MoS$_2$ monolayer was an electron acceptor, while at higher concentrations, it was an electron donor. Qui et al. [43] investigated the effects of S and Mo monovacancies on the SBH at the Au/MoS$_2$ junction, and found that there was a minor increase by ~5% in the SBH due to S-vacancies at defect concentration $n_{V_s}$~2%, while the effect of Mo-vacancies was significantly stronger than that of S-vacancies, resulting in the disappearance of SBH. Su et al. [44] confirmed that the SBH could be eliminated when Mo monovacancies were present in the MoS$_2$ monolayer at a critical concentration, while S monovacancies increase the SBH of Pt/MoS$_2$ interface [45]. It was revealed that due to the presence of Mo-vacancies, chemical bonds were formed between the monolayer and its underlying substrate, resulting in a transformation of Au/MoS$_2$ junction from a Schottky contact to an Ohmic contact [44]. Fang et al. [46] found that the SBH increased in MoS$_2$ when contacted with Mg, Al, In, and Au, while reduced in defective MoS$_2$ when contacted with Cu, Ag, and Pd.

Experimental studies [19, 41] have shown that MoS$_2$ samples may contain different types of defects, for example, S monovacancy, S divacancy, antisite defect. For the same type of defect, it may take different locations, for example, at the top sublayer or the bottom sublayer. Also, the defect may have varying defect densities. An interesting question is: how do defect type, location, and density in the MoS$_2$ layer affect the SBH of a metal/MoS$_2$ heterojunction? To answer this question, we choose the Au(111)/MoS$_2$ heterojunction to systematically examine these effects by leveraging our expertise in first-principles calculations. The Au(111) substrate is chosen because of its well-known chemical inertness, strong electronegativity, and stability [23]. Three types of point defects are chosen: S monovacancies, S divacancies, and Mo$_S$ antisite defects (in which an S atom is substituted by a Mo atom). In addition, the same defects located on the outer sublayer and inner sublayer of MoS$_2$ are also considered and compared. Finally, the effect of defect density on the SBH is also examined. Ultimately, we would like to find out whether these factors can explain the broad variation in the experimentally measured SBH values and propose possible strategies to control the SBH.

Two different first-principles-based methods can be used to calculate the SBH in the MoS$_2$/MoS$_2$ contact. The first method is based on projection of electronic band structure of MoS$_2$ layer taken from the Au/MoS$_2$ heterojunction on the band structure of the entire junction while the second method is based on the Schottky-Mott (SM) rule [15], which requires the attainment of the work function ($W_{Au}$) for Au substrate, electronic affinity energy (EAE) for MoS$_2$, and the step in electrostatic (Hartree) potential of Au/MoS$_2$ contact. Since the calculation results from these two methods are often different, in this study, we employ both methods and compare their predictions and assess their reliability and accuracy.

Our first-principles calculations show that both calculation methods predict the same trend for SBH, and with proper treatments, the two methods can predict nearly the same results. Our calculations also show that the SBH of the Au/MoS$_2$ contact is affected by defect type, location, and density in MoS$_2$ monolayer. More specifically, the SBH in the Au/MoS$_2$ contact with the defective MoS$_2$ monolayers is universally higher than that in the defect-free layer. Among the defects considered, Mo$_S$ antisite defect and S divacancy significantly increase the SBH, while the SBH is only weakly affected by S monovacancy. Moreover, the defects in the inner sublayer have more influence on SBH than those in the outer sublayer. Finally, an increase in the defect density noticeably increases the SBH. Our study suggests that the reported variations in the experimentally measured SBH for Au/MoS$_2$ contact can to a certain degree be accounted by the variations in the type, location, and density of point defects in MoS$_2$ monolayer. However, the predicted SBH values are ubiquitously higher than the experimentally measured values. We suggest that the lower SBH values observed in experiments may be due to the difference in experimental samples. The present study indicates that the value of SBH can be altered via defect engineering in the MoS$_2$ layer. Our findings provide a guide for tuning the SBH in the Au/MoS$_2$ heterojunctions.

## 2. Computational model

In our DFT calculations, we selected three different types of point defects with a relatively low formation energy. (1) S monovacancy: Its formation energy is $E_S^{vac}$=1.55 eV [22,47] in Mo-rich limit (deficit of S-atoms), and $E_S^{vac}$=2.81 eV [22,24,47] in S-rich limit (deficit of Mo-atoms). (2) S divacancy: Its formation energy is $E_{2S}^{vac}$=3.2 eV [30,48] in Mo-rich limit and $E_{2S}^{vac}$=5.44 eV in S-rich limit [22]. Since the formation energy of a S divacancy is approximately twice of that of an S monovacancy, this implies that S monovacancies do not have a strong tendency to merge, which is in contrast to graphene, where divacancies are energetically more favorable than monovacancies [30]. (3) Mo$_S$ antisite defect (an Mo-atom substitutes an S-atom): Its formation energy is $E_{Mo \rightarrow S}^{sub}$=4.2 eV [47] in S-rich limit, and $E_{Mo \rightarrow S}^{sub}$=6.2 eV [47] in Mo-rich limit. Due to the relatively higher formation energy, the antisite defects are likely to form at high temperatures [26].

Since the formation energy of a Mo monovacancy is $E_{Mo}^{vac}$=7.2 eV [47] and $E_{Mo}^{vac}$=8.2 eV [7] in Mo-rich limit, and $E_{Mo}^{vac}$=4.9 eV [30] in S-rich limit, and once a Mo monovacancy is formed, its nearby S-atoms have a strong tendency to leave behind vacancies since the S monovacancy formation energy around an Mo monovacancy is only ~1.1 eV even under S-rich condition [30], therefore, Mo monovacancies are not observed experimentally alone, but as clusters of vacancies. Yet the clusters of vacancies, which appear via merging of S and Mo monovacancies, such as: V(MoS$_2$) $E_{MoS_2}^{vac}$=8.2 eV [30] and V(MoS$_3$) $E_{MoS_3}^{vac}$=4.5 eV [30] in S-rich limit, are rather unstable especially when MoS$_2$ layer is supported by Au substrate [22]. This is consistent with the experimental observations, where S monovacancies are frequently observed in all samples, but Mo monovacancies were only occasionally found [30]. We note that when MoS$_2$ monolayer interacts with the Au substrate, the formation energies of these point defects are slightly higher [22]. For example, the formation energy of an S monovacancy increases by 7% ($E_S^{vac}$=2.81 eV -> $E_S^{vac}$=3.0 eV), and an S monovacancy in the top sublayer is slightly more stable than in the bottom sublayer.

As the first step, we constructed and optimized the Au(111)/MoS$_2$ samples with a vacuum thickness of 20 Å, constraining the lattice constants. The following six Au/MoS$_2$ samples were constructed (see Figure 1): (1) Defect-free (PF) MoS$_2$, (2) MoS$_2$ with an S monovacancy in the top sublayer (VT), (3) MoS$_2$ with an S monovacancy in the bottom sublayer (VB), (4) MoS$_2$ with an S divacancy (DV), (5) MoS$_2$ with an Mo$_S$ antisite defect at the top sublayer (AST), and (6) MoS$_2$ with an Mo$_S$ antisite defect in the bottom sublayer (ASB). The top and side views of the defect-free sample are shown in Figure 1 (a), while the samples with a VT and AST of MoS$_2$ layer are shown in Figure 1 (b) and (c), respectively. The defect sites are indicated by red arrows. The samples with an ASB and a DV of MoS$_2$ layer are shown in Figures 1 (d) and (e), respectively. Planar charge density distribution around the point defects in the Au/MoS$_2$ samples are shown at the bottom panel in Figure 1.

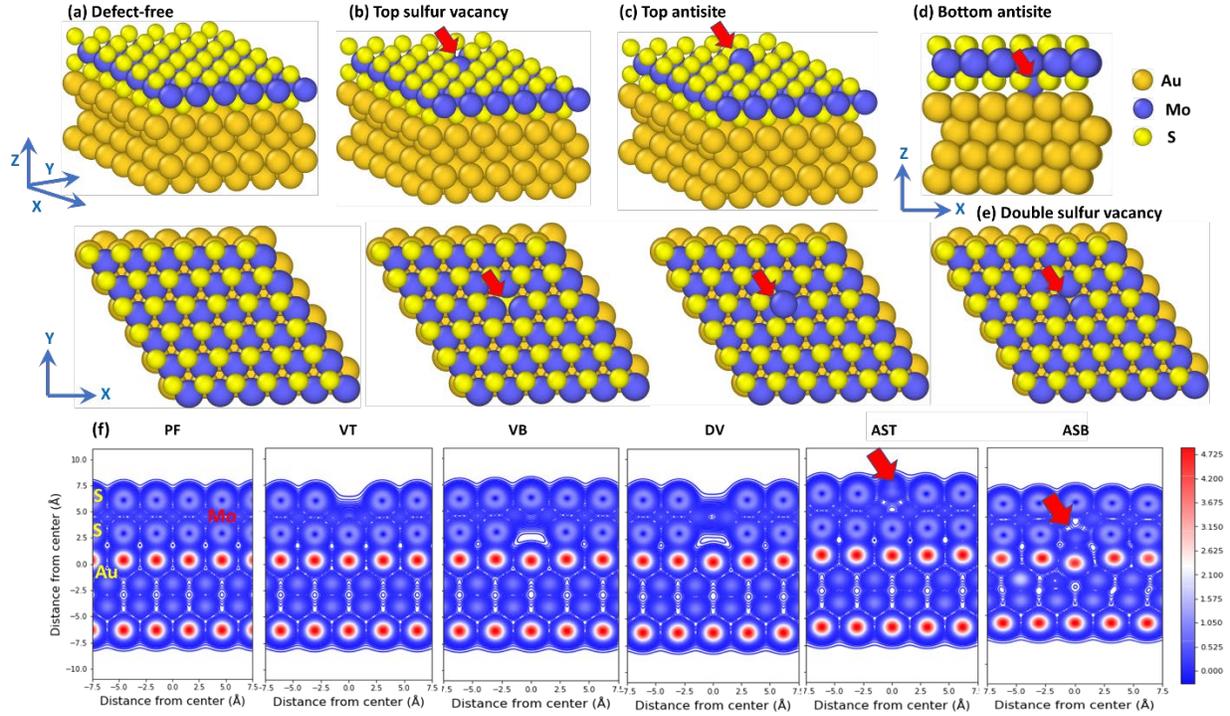

Figure 1: Side and top views of the Au (111)/MoS$_2$ 6x6x4 samples. (a) defect-free sample (PF), (b) sample with an S monovacancy in the top sublayer (VT), (c) sample with an antisite defect in the top sublayer (AST), and (d) sample with an antisite defect in the bottom sublayer (ASB), and (e) sample with a double S vacancy (DV). Mo atoms marked by blue, small yellow spheres correspond to S-atoms, and large ones to Au-atoms. (f) Planar charge density distribution around the point defects in the Au/MoS$_2$ 5x5x4 samples (bottom panel): PF: Defect-free MoS$_2$, (2) VT: MoS$_2$ with an S monovacancy in the top sublayer, (3) VB: MoS$_2$ with an S monovacancy in the bottom sublayer, (4) DV: MoS$_2$ with a double S vacancy, (5) AST: MoS$_2$ with an antisite defect at the top sublayer, and (6) ASB: MoS$_2$ with an antisite defect in the bottom sublayer. Color bar indicates charge density values. Red arrow indicates the defect position.

To study the effect of defect density on the SBH, we varied the lateral size of the computational cell: The supercells of 3x3, 4x4, 5x5 and 6x6 lattice unit cells of MoS$_2$ monolayer accommodated on Au (111) substrate were constructed. The defect density (and vacancy concentration) per unit area in the constructed samples is given in Table 1. For 3x3 MoS$_2$ supercell, we built Au (111) substrates containing 4, 5 and 6 Au layers; but found that the difference in the obtained SBH values was rather minor, thus for the remaining samples with 4x4, 5x5 and 6x6 supercells, we constructed Au (111) substrate with 4 layers. The supercells were relaxed, while the positions of atoms in the two bottom layers of the Au (111) substrate were constrained and the positions of the remaining Au atoms were relaxed.

Table 1: Defect density per unit area for the constructed samples.

| Au/MoS$_2$ sample size | Defect density per unit area (1/Å$^2$) | Defect concentration |
|---|---|---|
| 6x6x4 | 0.003 | 1% |
| 5x5x4 | 0.005 | 2% |
| 4x4x4 | 0.008 | 3% |
| 3x3x4 | 0.014 | 6% |

Periodic boundary conditions were applied along all the directions, while a vacuum layer with the thickness of ~20 Å was added as a padding along the Z-direction (normal to the Au(111) surface, see Figure 1) to avoid interactions due to periodic boundary conditions. Since there was a small lattice mismatch along the lateral (X, Y) directions between the lattice constants of primitive unit cells of $MoS_2$ monolayer and that of Au(111) surface, the Au(111) substrate was deformed to eliminate the mismatch. This is a common practice [43,49], which permits the application of periodic boundary conditions in DFT calculations. The physical basis for this treatment is that a minor tensile deformation of the metallic substrate by a few percent only leads to a minor change in its electronic band structure and work function. The geometry of the constructed samples was optimized by DFT method using conjugate-gradient optimization.

Two first-principles-based methods, the projection of electronic band structure and the SM rule, have been frequently used to calculate SBH. We use both methods to calculate SBH and compare and assess their reliability and accuracy. Below, we briefly discuss these two methods.

### The method based on projection of electronic band structure

In this method, the SBH value is obtained by identifying the position of conduction band minimum (CBM) of the contact $MoS_2$ layer among the bands of the Au(111) /$MoS_2$ heterojunction. The value of SBH is the distance from the Fermi level to the identified CBM [18,37]. Hence, to calculate SBH, one needs to obtain the electronic band structure of the Au(111)/$MoS_2$ heterojunction, and that of the contact $MoS_2$ monolayer taken from the Au(111)/$MoS_2$ heterojunction. Since when a free-standing $MoS_2$ layer is accommodated on a substrate, its geometry, and therefore its electronic band structure may be changed, therefore, to calculate its electronic band structure, we take the $MoS_2$ layer from the Au(111)/$MoS_2$ heterojunction by freezing its geometry. By using the frozen $MoS_2$, one can obtain its CBM accurately (see the red-colored band structure in Figure 2(a)).

Next, the electronic band structure of contact $MoS_2$ layer is projected onto the electronic band structure of Au(111)/$MoS_2$ heterojunction (see Figure 2 (a)). When the superimposed electronic bands align, one can identify the minimum of the electronic band of the Au(111)/$MoS_2$ heterojunction by looking for a band that overlaps to a greater extent with the CBM band of the frozen $MoS_2$ monolayer. In Figure 2 (b), the red-colored bottom conduction band matches with one of the bands of the Au/$MoS_2$ heterojunction (see the blue line in Figure 2 (b)). The distance from the Fermi level to the identified minimum (which is indicated by the red arrow in Figure 2 (b)) is equal to the SBH of the of Au(111)/$MoS_2$ heterojunction [11,18,40].

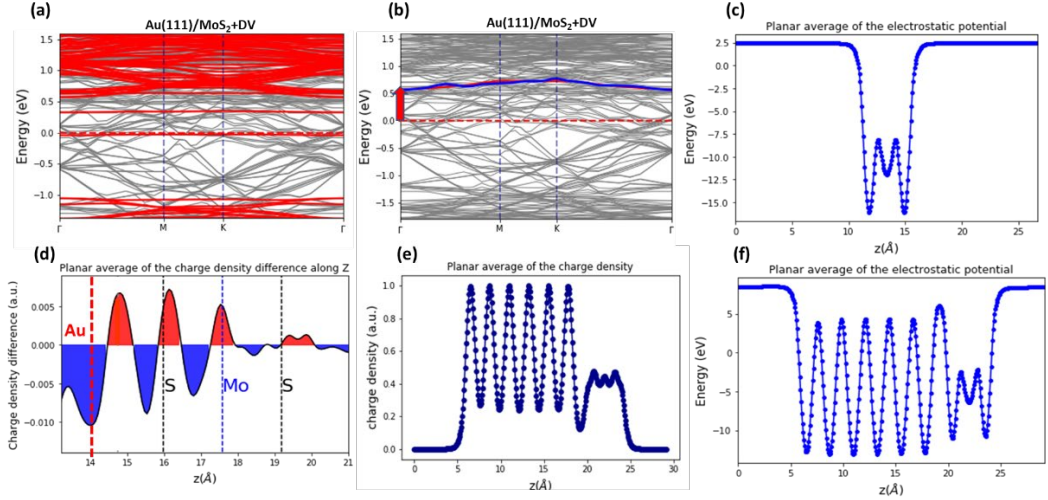

Figure 2: (a) Electronic band structure of the MoS$_2$ monolayer with a double sulfur vacancy (red bands) superimposed over the band structure of Au/MoS$_2$ junction (grey bands). (b) The superimposed CBM band of MoS$_2$ monolayer containing a double sulfur vacancy (red band) is matched in a high accuracy with one of the Au/MoS$_2$ junction bands (blue band). The distance from the minimum of the matched band to the Fermi level indicates the SBH value (shown by arrow at Γ-point). (c) Planar average of charge density of the MoS$_2$ monolayer with a double sulfur vacancy. (d) Planar average of charge density difference for Au/MoS$_2$ junction with a double sulfur vacancy. Red color indicates the charge accumulation regions and blue the charge depletion regions. The dashed lines indicate the average Z-position of Au-atoms at the top surface layer, and S-atoms in the top and bottom MoS$_2$ sublayers, as well as Mo-atoms in the middle sublayer. (e) Planar average of charge density of the Au/MoS$_2$ junction. (f) Planar average of Hartree potential of the Au/MoS$_2$ junction a double sulfur vacancy in the MoS$_2$ monolayer. The Z-axis is normal to the Au/MoS$_2$ contact plane and the plane average is calculated over [XY] planes along the sample. The plots are for the Au/MoS$_2$ 3x3x6 sample.

An important restriction to use this approach is that the monolayer-substrate interactions must be weak enough so that the weak interactions should only perturb the band structure of MoS$_2$ to a small extent. In the case of the Au/MoS$_2$ heterojunction, since the interfacial bonding is attributable to van der Waals interactions [18,40], this method is applicable to the Au/MoS$_2$ junction. Therefore, we applied this method for both defect-free and defective MoS$_2$ monolayer. It is noted that point defects introduce new occupied defect states below and unoccupied states above the Fermi level in the band gap of MoS$_2$ monolayer. Since the vacancy produces localized states [50], we used the CBM position to obtain the SBH.

In addition to the electronic band structure, partial density of states (pDOS) is a convenient way to illustrate the effect of point defects in the MoS$_2$ layer on the electronic properties of Au/MoS$_2$ junction (see Figures 4-6). The pDOS is calculated separately for the Mo- and S-atoms of the contact layer as an average over all the atoms and their corresponding orbitals (five 4d-orbitals for Mo-atoms and three 3p-orbitals for S-atoms). The position of CBM cannot be identified from the pDOS plots with high accuracy since the band edge shape in pDOS plot is often fuzzy. The exact position was taken by using the method based on the projected electronic band structure.

## The method based on the SM rule

Another commonly used method to calculate the SBH is based on the SM rule [15]. According to this rule, the value of SBH between a metal/semiconductor junction is proportional to the difference of metal work function, $W_m$, and the semiconductor electron affinity energy, $\chi$: $\Phi = W_m - \chi$. For a metal, which is in our case Au(111) substrate, the work function is defined as the difference between its vacuum energy level and the Fermi energy. We obtained $W_{Au}$=5.1 eV from our DFT calculations with PBE XC-functional, and $W_{Au}$=5.27 eV with PBE XC-functional and DFT-D2 van der Waals correction. It is noted that the calculated values are slightly lower than previously reported values of 5.13 eV and 5.3 eV [11,18] since we deformed the Au(111) sample to enable the application of periodical boundary conditions.

The electron affinity energy (EAE), denoted as $\chi_{MoS_2}$, is calculated as the difference between the vacuum energy level (obtained as an asymptotic value of planar averaged electrostatic Hartree potential, which is taken sufficiently far off the monolayer, see Figure 2 (c)) and the energy level of the conduction band minimum, which is identified by using the calculated electronic band structure of the MoS₂ layer. In our case, the $\chi_{MoS_2}$ varies within a certain range around $\chi_{MoS_2} = 4.2\ eV$ for defective monolayer (see Figure 7(b) and Tables S1-S4 in Supplementary Materials).

To account for the interaction between the MoS₂ monolayer and the underlying metallic substrate and the corresponding change in the work function of the metal in the presence of MoS₂ layer, the SM rule must be modified. When the MoS₂ monolayer and Au substrate is integrated into the Au/MoS₂ junction, the equalization of the Fermi levels results in the charge transfer from the metal to the MoS₂ monolayer (see Figure 2 (d), where the charge accumulation and depletion zones at the Au (111)/MoS₂ junction are exemplified), which alters the SBH. The charge transfer and its redistribution at the Au/MoS₂ junction results in the potential step, $\Delta V$, given by $\Delta V = \frac{e^2}{A} \iiint z\Delta n(x,y,z)dxdydz$, where $A$ is the contact area (measured within the [X,Y] plane), and $\Delta n(x,y,z) = n_{Au/MoS_2}(x,y,z) - n_{Au}(x,y,z) - n_{MoS_2}(x,y,z)$ is the difference between the electronic density of Au/MoS₂ junction, $n_{Au/MoS_2}$, (which is illustrated in Figure 2(e) for the Au/MoS₂ junction containing double S-vacancies) and the electronic density of Au substrate, $n_{Au}(x,y,z)$ and that of MoS₂ monolayer, $n_{MoS_2}(x,y,z)$. According to the modified SM rule, which includes the effect of the interface potential step, the SBH value is given by: $\Phi_{Au/MoS_2} = W_m - \chi_{MoS_2} - \Delta V$ [18,40]. The interface potential step is attributed to the reduction in the metal work function due to its contact with the MoS₂ monolayer. The change in the work function $W_m$ is a combined effect due to the rehybridization of d-orbitals of Au-atoms [13], polarization of the metal electrons induced by the MoS₂ monolayer [51], the "pushback" effect (the displacement of surface electron density around the metallic substrate into the metal by the MoS₂ monolayer) due to the exchange (Pauli) repulsion at the interface, which is the main contribution to the interface potential step in the weakly interacting regime [40,52,53], the presence of interface dipole moment [18] and the surface relaxation of metallic substrate [37,40].

The potential step at the interface can be calculated either by using the planar average electronic charge density along the z-direction, $n(z)$, or by using the plane-averaged Hartree potential defined as $V(z) = \frac{e^2}{A} \iint z\Delta n(x,y,z)dxdy$. According to Farmanbar et al. [40], the potential step can be obtained by inspecting the asymptotic values of $V(z)$ for the Au/MoS₂ junction in the vacuum, which are typically attained within a few Å from the metallic surface at the bottom and the MoS₂ layer at the top (see Figure 2 (e), where the plane-averaged Hartree potential is shown for the Au(111)/MoS₂ junction with double S-vacancies). Thus, one can calculate the value of $\Delta V$ as the difference of $V(z)$ taken between two points located at sufficiently large distance deep in the vacuum (at the points where electrostatic potential $V(z)$ converges to constant values). Since the periodic boundary conditions are applied in the DFT calculations, one needs to use dipole corrections along the z-axis to obtain the well-defined potential step in $V(z)$.

## Comparison of the two methods

To compare these two methods, we plot the calculated SBH values obtained from the method based on the band structure projection (see blue circles in Figure 3 (a-d)) and the method based on the SM rule (see red squares in Figure 3 (a-d)). The results from these two methods show a remarkably similar trend between the SBH and defect type as shown in Figure 3 (a-d) (see also Tables S1-4 and Figure S5 in Supplementary materials). On average, the difference in the SBH values is ~3%, while the maximal difference is 7%. We note that the difference in the SBH values obtained by the two methods in this study are smaller than previously reported 0.68 eV and 0.60 eV [40].

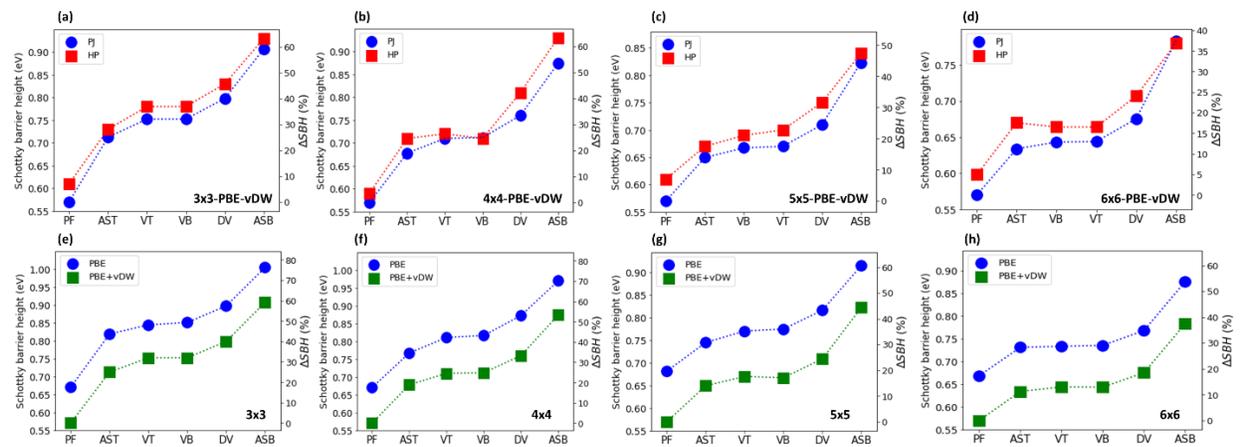

Figure 3: (a) Comparison of the SBH calculated with the method based on projection (PJ) of electronic band structure (blue circles) and the method based on the modified SM rule using Hartree potential (HP, red squares). On the left vertical axis, the SBH values for Au (111)/MoS$_2$ junction with PF, VT, VB, DV, AST, and ASB. The data is for the 3x3 (with defect density $n_d$=14x10$^{-3}$ Å$^{-2}$), 4x4 ($n_d$=8x10$^{-3}$ Å$^{-2}$) , 5x5 ($n_d$=5x10$^{-3}$ Å$^{-2}$) and 6x6 ($n_d$=3x10$^{-3}$ Å$^{-2}$) Au(111)/MoS$_2$ junctions. On the right axis, the relative increase of the SBH with respect to the defect free sample $\Delta SBH(\%) = 100 \times \left(\frac{SBH - SBH_0}{SBH_0}\right)$ (b, c) Comparison of the SBH values calculated with the PBE exchange-correlation (XC) potential (blue circles) and PBE-XC and van der Waals DFT-D2 corrections (green squares) for Au(111)/MoS$_2$ junctions. The defect density is the same as in (a).

## Details of DFT calculations

All our calculations were carried out by using density functional theory (DFT) with the generalized Perdew-Burke-Ernzerhof [54] and the projector-augmented wave (PAW) pseudopotential plane-wave method [55] for the core electrons as implemented in the Vienna ab initio simulation package (VASP) code [56]. For the PAW pseudopotentials, we included 5d$^{10}$6s$^1$, 4p$^6$d$^5$5s$^1$, and 3s$^2$p$^4$ as valence electrons for Au, Mo, and S, respectively. For DFT calculations, we used 6 × 6 × 1 Monkhorst–Pack [57] k-point grid for the geometry optimizations, and a plane-wave basis set with an energy cut-off of 520 eV was adopted. Good convergence was obtained with these parameters, and the total energy was converged to $10^{-7}$ eV per atom. The atomic samples were fully relaxed with a residual force of less than 0.02 eV/Å. Spin polarization was considered in this study. The energy minimization was performed using a conjugate-gradient algorithm to relax the ions into their instantaneous ground state. The DFT calculations were done with van der Waals corrections using Grimme's DFT-D2 approach as realized in the VASP [56]. Dipole corrections to the total energy were used along the direction normal to Au(111)/MoS$_2$ interface for all calculations [58]. To avoid spurious interactions between replica of the Au(111)/MoS$_2$ interface, a vacuum region of at least 20 Å is included along the same direction normal to the junction in the supercell.

We note that the application of van der Waals corrections not only leads to more accurate results, but it is crucial for Au substrate: In contrast to other more reactive metallic surface like Mo and Ti, where bonds are formed [13], the van der Waals nature of the Au–MoS$_2$ interaction is prevalent. Covalent bonds between Au and S atoms cannot be formed since the Au atom with one s-electron has fully occupied d-orbitals, and hence only weakly interacts with MoS$_2$ [40]. In Figure 3(e-h), we compare the SBH calculated with (green squares) and without (blue circles) van der Waals corrections. It is evident that application of van der Waals corrections systematically lowers the SBH values by ~15% (~0.1 eV).

## Beyond PBE functional: hybrid HSE XC-potential

There is an uncertainty in the calculated SBHs coming from using PBE functional. In our DFT calculations, we used the PBE functional, but it is well-known that it underestimates the band gap of MoS$_2$ since it does not take into account the many body effect among electrons, only partially accounts for electronic correlation, and neglects long-range exchange and subtle screening effects [24,59]. We obtained a direct band gap of $E_g$=1.7 eV for MoS$_2$ monolayer using PBE with DFT-D2 van der Waals corrections, which is in a good agreement with previous GGA calculations [11,20], while calculations based on the GW-quasiparticle approximation give $E_g$=2.8 eV [60,61] and application of hybrid HSE XC-potential results in $E_g$=2.2 eV [26]. Even though the electronic band gap for free-standing MoS$_2$ monolayer is not well known, the results obtained with HSE and GW-quasiparticle approximations are in excellent agreement with the experimentally measured optical band gap is $E_g$=2.9 eV [21,62]. It must be admitted that the electronic band gap is a fundamentally different from the optical gap, which is generally measured by photoluminescence experiments [63]. The optical band gap corresponds to the energy required to create an exciton, while the electronic band gap also requires the breaking of the exciton, and is thus higher due to the exciton binding energy. Exciton binding energies between 0.01- 0.5 eV have been reported [40]. A direct comparison of the PBE vs. GW-quasiparticle approximation is not truly fair, as the observed difference includes the exciton binding energy, obtained by using the GW-quasiparticle approximation.

We estimated the required corrections when the hybrid density XC-functional potential is applied. Hybrid functionals mix a fraction of the short-range part of the Hartree-Fock (HF) exchange interaction with the local functional. There is a range of hybrid functionals, among them, we selected the Heyd, Scuseria, and Ernzerhof (HSE) hybrid functional [64]. This hybrid density functional is based on a screened Coulomb potential for the exchange interaction which circumvents the bottleneck of calculating the exact (Hartree–Fock) exchange, especially for systems with metallic characteristics. The main reason for the selection is due to its high accuracy combined with its computational advantages for periodic systems [64]. Moreover, the conduction band in MoS$_2$ consists of d-orbitals and PBE functional has significant limitations in proper description of localized d-electron states. Therefore, we complement our DFT calculations with hybrid functional calculations for the band structures, Hartree potential and defect states.

# 3. Results and discussions

## Defect-free sample

First, we examined the effect of different types of point defects on the value of SBH. To investigate this effect, we used the Au(111)/MoS$_2$ junction with a defect-free MoS$_2$ monolayer as a reference, which was compared with the samples containing defects. In Figure 4 (a), we plot the pDOS of a defect-free free-standing MoS$_2$ monolayer, which was calculated as an average over 4d-orbitals of Mo-atoms and 3p orbitals of S-atoms. For comparison, the pDOS of the MoS$_2$ layer taken from the Au (111)/MoS$_2$ junction is shown in Figure 4 (b). It should be readily seen that the rearrangement of atomic position in the contact layer due to its interaction with the Au substrate changes the overall shape of pDOS, but the band gap and the location of the CBM are nearly the same.

The pDOS of the Au(111)/MoS$_2$ sample and the corresponding electronic band structure are shown in Figure 4 (c) and Figure 4 (d), respectively. In addition, we project the electronic band structure of the contact MoS$_2$ layer onto the Au (111)/MoS2 sample band structure (see the red-colored electronic bands in Figure 4(d)). The electronic band structure and pDOS are changed noticeably due to the interaction of the MoS$_2$ layer with its underlying Au(111) substrate.

The mid-gap states appear in the band gap of MoS$_2$ monolayer as shown in Figure 4 (c). Direct orbital hybridization occurs between Au- and S-atoms at the Au/MoS$_2$ interface due to the overlap of their wave functions, while S-atoms mediate indirect orbital hybridization between Au- and Mo-atoms, resulting in formation of mid-gap states [11]. The Fermi level at the interface, which determines the SBH, is now governed by the charge transfer and filling of the mid-gap states. Although the density of mid-gap states is somewhat low, it is sufficient to pin the Fermi level above the middle of MoS$_2$ band gap, as in an n-type contact [17,22]. Fermi pinning sets the Fermi level close to the MoS$_2$ CBM, preventing from reaching it.

The position of MoS$_2$ CBM in the pDOS of the Au(111)/MoS$_2$ contact is indicated in Figure 4 (c). If one compares the pDOS of the free-standing MoS$_2$ monolayer in Figure 4(a, b) with the pDOS of the Au(111)/MoS2 sample, it is apparent that the position of the CBM edge, which determines the SBH value, is shifted closer to the Fermi level, eventually setting the SBH value. To accurately pinpoint the CBM edge location, we use the projection method as shown in Figure 4 (d), which is consistent with the pDOS estimate. We note that the CBM located at the Γ point in the reciprocal space of the $\sqrt{3}x\sqrt{3}$ superlattice of single-layer MoS$_2$ is in accordance with previous reports [8,18,40]. The SBH value for the Au(111)/MoS$_2$ contact sample with defect-free monolayer obtained with the Hartree potential method is reported in Supplementary materials.

## S Monovacancies

Next, VT and VB defects are introduced in the MoS$_2$ monolayer (see Figure 4 (e)). An introduction of S monovacancy creates dangling bonds in the neighboring Mo-atoms, which lead to a defect state in the band gap positioned close to the bottom of conduction band (see also the distinct peak in the pDOS of a free-standing MoS$_2$ monolayer in Figure 4 (f)). The new defective state is mainly due to the dominant 4d-states of Mo-atoms with only a small mixture of 3p states of S-atoms.

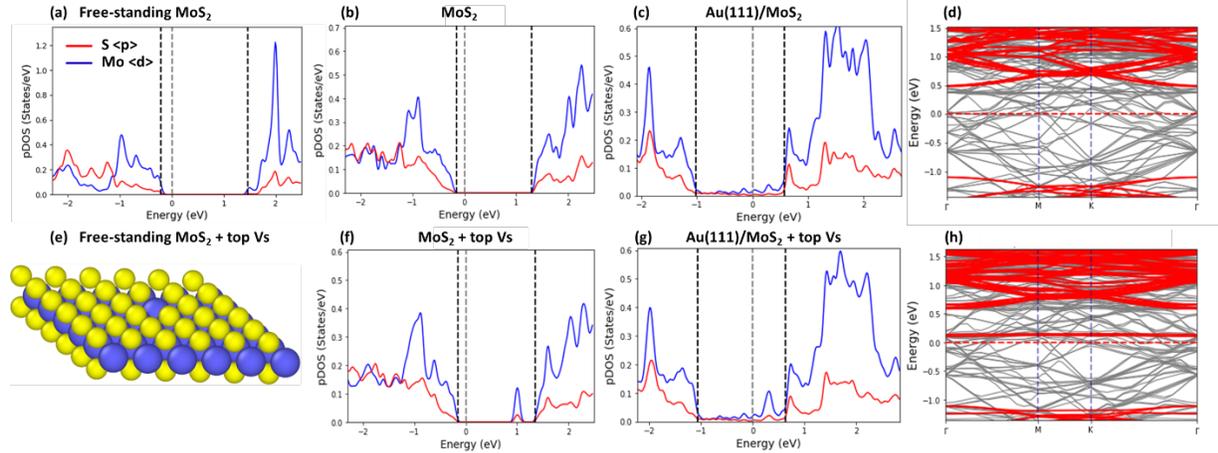

Figure 4: Partial density of states (pDOS) of a free-standing defect-free MoS$_2$ layer (a) and the contact MoS$_2$ layer (b) taken from the corresponding Au/MoS$_2$ junction. pDOS of the Au/MoS$_2$ sample is shown in (c). pDOS calculated as an average over d-orbitals of Mo-atoms indicated by blue, and over p-orbitals of S-atoms indicated by red. The valence band maximum (VBM), Fermi level and conduction band minimum (CBM), obtained with the PJ method are shown by dashed lines. (d) Electronic band structure of the contact layer (red bands) superimposed over the band structure of Au/MoS2 sample (grey bands). (e) Contact MoS$_2$ layer containing a top sulfur monovacancy taken from the respective Au/MoS$_2$ sample. (f, g) pDOS of the contact layer (f) and Au/MoS$_2$ sample (g). (h) Electronic band structure of the contact layer with a VT defect (red bands) superimposed over the band structure of the Au/MoS$_2$ sample (grey bands). The sample size is 6x6x4 with PBE XC+ van der Waals DFT-D2 corrections.

The pDOS for atoms in the MoS$_2$ layer taken from the Au(111)/MoS$_2$ sample is fairly similar to that of the free-standing MoS$_2$ layer (see Figure 4 (f) for VT defect and Figure 5(b) for the VB defect). However, due to the interaction of MoS$_2$ layer with its underlying substrate, the position of the peak in the band gap corresponding to the defect state shifts closer to the bottom of conduction band.

In the pDOS for Mo- and S-atoms of the Au(111)/MoS$_2$ sample with the MoS$_2$ layer containing S monovacancies (see Figure 4 (g) for the VT defect and Figure 5 (c) for the VB defect), new states (with low density amplitude) appear in the band gap due to the mixing and hybridization of S-atom orbitals, and some extent Mo-atom orbitals with the orbitals of surface Au-atoms. The height of the vacancy-related peak somewhat diminishes, while its width broadens. We used the projection method based on the electronic band structure of the Au(111)/MoS$_2$ samples to identify the CBM positions (see Figure 4 (h) for the VT defect and Figure 5 (d) for the VB defect, respectively), and found that the SBH increases in the range from ~10% to ~30% due to the S monovacancies. It should be noted that the SBH also depends upon the number of S monovacancies per unit area, which will be discussed later.

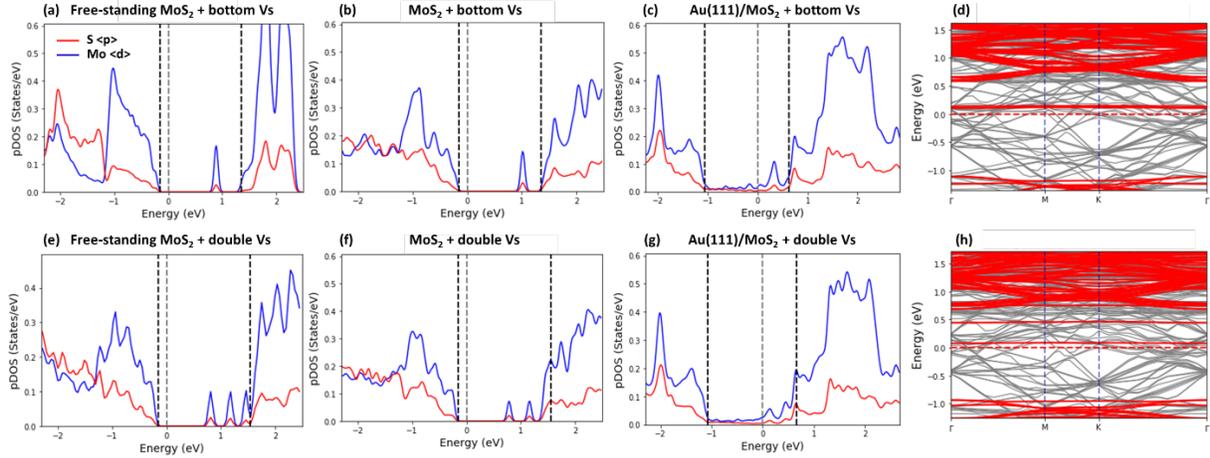

Figure 5: (a-c) pDOS of a free-standing MoS$_2$ layer with a single sulfur vacancy (a) and the contact MoS$_2$ layer with a VB defect (b) taken from the respective Au/MoS$_2$ sample. pDOS of the Au/MoS$_2$ sample is shown in (c). (d) Electronic band structure of the contact layer (red bands) superimposed over the band structure of Au/MoS2 sample (grey bands). (e-g) pDOS of a free-standing MoS$_2$ layer with a DV defect (e) and the contact MoS$_2$ layer with a DV defect (f) taken from the corresponding Au/MoS$_2$ sample. pDOS of the Au/MoS$_2$ sample is shown in (g). (h) Electronic band structure of the contact layer (red bands) superimposed over the band structure of Au/MoS$_2$ sample (grey bands). The sample size is 6x6x4, for PBE XC+ van der Waals DFT-D2 corrections.

## S divacancies

Next, we calculated the pDOS for S- and Mo-atoms of a free-standing MoS$_2$ monolayer with S divacancies, which were created by removing S-atoms from both the top and bottom sublayers of MoS$_2$ layer as shown in Figure 1 (e). The pDOS results are shown in Figure 5 (e). It is seen that S divacancies result in the three distinct peaks in the band gap located above the Fermi level, with one of them being near the bottom of the conduction band.

The pDOS for S- and Mo-atoms of the MoS$_2$ monolayer with S divacancies taken from the Au(111)/MoS$_2$ sample is shown on Figure 5 (f). The rearrangement in the atomic positions of the defective MoS$_2$ due to its interaction with Au(111) substrate modifies the pDOS, especially the shape of the peak in the proximity to the bottom of the conduction band.

In Figure 5 (g), we plot the pDOS for S- and Mo-atoms of the MoS$_2$ layer with S divacancies accommodated on Au(111) surface. The interaction of the defective MoS$_2$ layer with the Au substrate significantly changes its pDOS. As can be seen in Figure 5 (g), new states, with a low-density amplitude, appear in the band gap around the two divacancy-related peaks. The peaks merge to some extent, forming a double hump shape, while the third peak merges with the bottom of the conduction band.

The applications of the method based on projection of electronic band structure (see Figure 5 (h)) and the method based on the SM rule show that in the presence of S divacancies, the SBH increases by ~20%-40% as compared to that of the Au (111)/MoS$_2$ contact sample with a defect-free monolayer. The effect of S divacancy is almost twice as larger as that of S monovacancy, and thus can be approximately considered as a linear superposition of two monovacancies.

## Antisite defects

The effect of antisite defects introduced in the top (AST) or bottom (ASB) sublayer of MoS$_2$ layer (see Figure 1 (c, d)) on the pDOS, band structure and SBH of the Au(111)/MoS$_2$ sample are illustrated in Figure 6. It is seen that antisite defects markedly change the pDOS of a free-standing MoS$_2$ monolayer (see Figure 6 (e)). Five localized defect states occur within the band gap of the free-standing MoS$_2$: Two states below the Fermi level and three above, with one being in the vicinity of the bottom of the conduction band (see Figure 6 (a)).

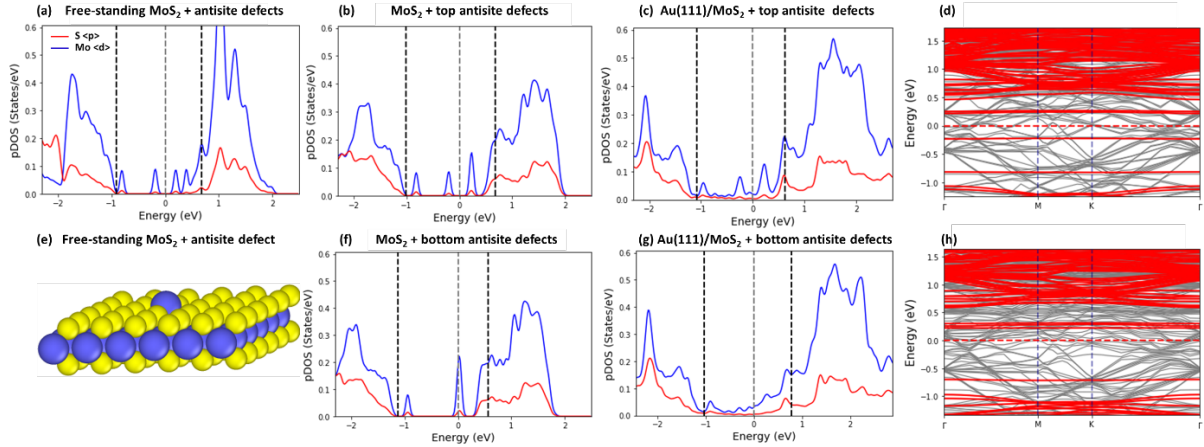

Figure 6: (a-c) pDOS of a free-standing MoS$_2$ layer with a single antisite defect (a) and the contact MoS$_2$ layer with an AST defect (b) taken from the respective Au/MoS$_2$ sample. pDOS of the Au/MoS$_2$ sample is shown in (c). (d) Electronic band structure of the contact layer (red bands) superimposed over the band structure of Au/MoS2 sample (grey bands). (e) The geometry of the contact layer with an AST defect (f). pDOS of the contact MoS$_2$ layer with an ASB defect (f) taken from the corresponding Au/MoS$_2$ sample. pDOS of the Au/MoS$_2$ sample is shown in (g). (h) Electronic band structure of the contact layer (red bands) superimposed over the band structure of Au/MoS$_2$ sample (grey bands). The sample size is 6x6x4 with PBE XC + van der Waals corrections.

When the MoS$_2$ layer is placed on Au(111) substrate, its electronic structure changes significantly as the result of its interaction with the underlying substrate. In Figure 6 (b), we plot the pDOS for atoms of MoS$_2$ layer with an AST defect taken from the Au (111)/MoS$_2$ contact (see also Figure 6 (f) for MoS$_2$ with an ASB defect). The geometry and the corresponding pDOS of the contact MoS$_2$ layer are markedly modified by the underlying substrate: The two peaks of a free-standing layer located above the Fermi level now merge into a single peak for the MoS$_2$ with an AST defect (see also the corresponding band structure in Figure 6 (h)). In contrast, three peaks located near the Fermi level now merge into one for the MoS$_2$ with an ASB defect (see Figure 6 (f) and the corresponding band structure in Figure 6 (h)). Contrary to the monovacancy defects, the difference in the pDOS between the AST and ASB defect in the MoS$_2$ monolayer is considerably larger.

Even more revealing is the change in the pDOS for S- and Mo-atoms of the MoS$_2$ monolayer with the ASB defects. As can be seen in Figure 6 (c), in the case of MoS$_2$ layer containing AST defects, the two defect states located above the Fermi level merge into one, while many additional states appear around it within the band gap. However, the shape of pDOS resembles that of the contact monolayer (or the free-standing monolayer). In the case of MoS$_2$ layer containing the ASB defects, the changes in the pDOS are substantial as compared with the contact (or free-standing) MoS$_2$ layer (see Figure 6 (g)). The different defect-related peaks merge with the new states within the band gap and form a broad continuum. This indicates that the interaction between the MoS$_2$ layer with the ASB defects is stronger than that with the AST defects. We calculated and compared the relative changes in the binding energy and the interfacial distance between the defective MoS$_2$ monolayer and its defect-free counterpart and found that the ASB defects have the strongest effect than the AST ones (see Figure S7 in Supplementary materials).

The methods based on the projection of electronic band structure (see Figure 6 (d) and Figure 6 (h)) and on the SM rule were applied to calculate SBH for MoS$_2$ layer with antisite defects. It was found that the presence of AST defects increases SBH in the range from ~10 to 25% (the SBH increases in direct proportion to antisite density). However, when ASB defects are present, the effect on the SBH is much more profound: The increase in SBH is in the range of ~ 40 to ~60% according to the number of the ASB defects per unit area.

## Comparison of different point defects

In Figure 7 (a), we summarize the obtained SBH results for Au/MoS$_2$ samples with a defect-free MoS$_2$ monolayer, as well as MoS$_2$ monolayer with VT, VB, DV, AST, and ASB defects. The impact of ASB and DV is the strongest, and that of VT and VB defects is in the middle, while that of AST defects is the weakest (see also Table S5 in Supplementary Materials).

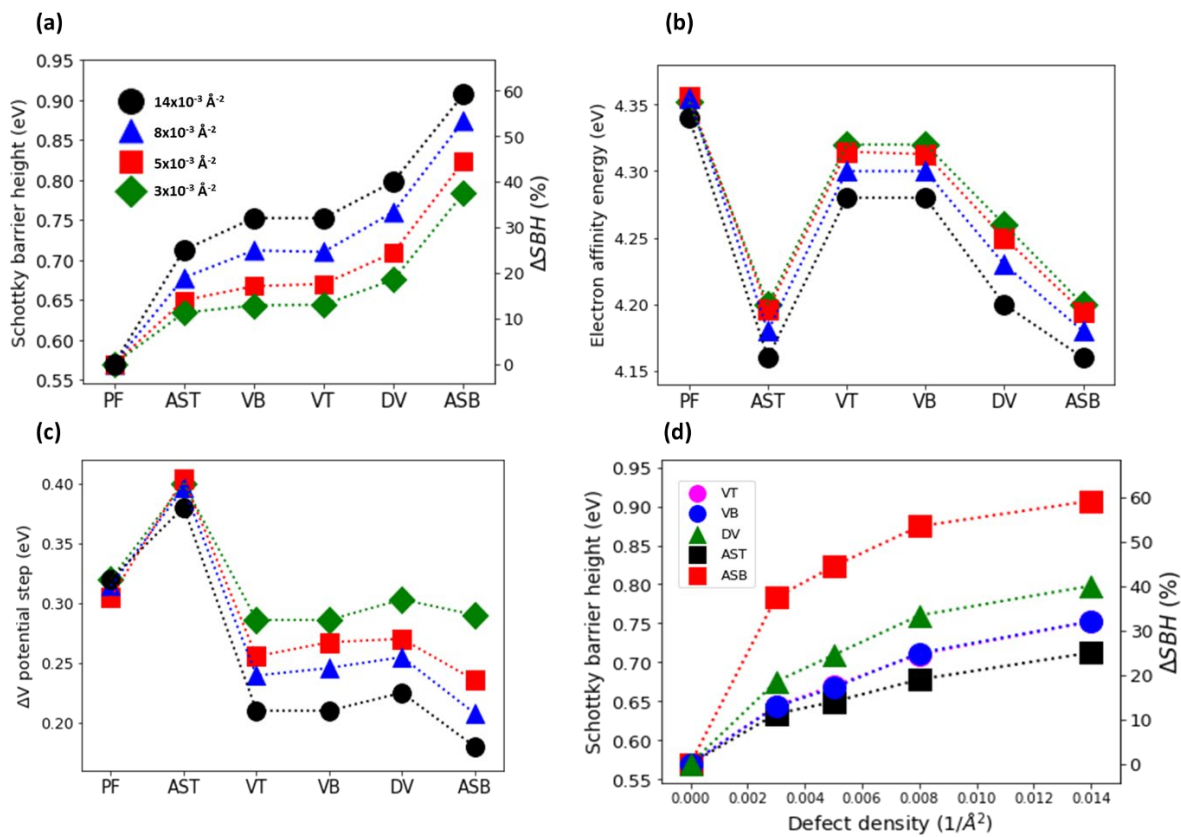

Figure 7: (a) The effect of point defects on SBH. On the left vertical axis, the SBH values for Au (111)/MoS$_2$ contact with a defect-free monolayer (PF) and a monolayer containing VT, VB, DV, AST, ASB defects. The data is for 3x3 samples with 6 Au layers (black circles), and 5x5 (red squares), 4x4 (blue triangles), and 6x6 (green diamonds) Au/MoS$_2$ samples with 4 Au layers. The corresponding defect densities are indicated. The right axis shows the relative increase of the SBH with respect to the defect free sample, that is, $\Delta SBH(\%) = 100 * \left(\frac{SBH - SBH_0}{SBH_0}\right)$. (b, c) Electron affinity energy (b) and potential step (c) of defect-free and defective MoS$_2$ monolayer for Au (111)/MoS$_2$ contact. (d) The effect of defect concentration on the SBH values for Au/MoS$_2$ sample with a MoS$_2$ monolayer containing VT (blue circles), VB (magenta circles), DV (green triangles), AST (black squares) and ASB (red squares) defects. The SBH value of the defect-free sample is given for comparison. The right axis shows the relative increase of the SBH with respect to the defect free sample. The DFT calculations with DT2 van der Waals (vDW) corrections are used. The SBH values are obtained based on projection of electronic band structure.

The value of SBH is calculated according to SBH = $W_{Au} - \chi_{MoS_2} - \Delta V$, where $W_{Au}$ is the work function of Au (111), $\chi_{MoS_2}$ is the electron affinity energy, and $\Delta V$ is the step in the Hartree potential, which represents the reduction of the work function of Au substrate in contact with MoS$_2$ layer. Since $W_{Au}$ of the substrate is fixed, the SBH increases when the value of $\chi_{MoS_2}$ for MoS$_2$ monolayer decreases. As can be seen from Figure 7 (b), the introduced point defects (especially the antisite defects and double vacancies) reduce noticeably the $\chi_{MoS_2}$. Besides that, the SBH increases when the $\Delta V$ decreases. As can be seen in Figure 7 (c), the hosted point defects (except for the AST defects) reduce the ΔV. However, even for the AST defects, the effect of the reduction in the value of $\chi_{MoS_2}$ is stronger than that of an increase in the value of $\Delta V$, thus the overall result is a minor increase in the SBH for these defects.

An interesting question is why the defects reduce $\chi_{MoS_2}$? It is known that the electron affinity is the energy required to transfer an electron from the bottom of the conduction band to the vacuum level. The $\chi_{MoS_2}$ in MoS$_2$ monolayer is measured as the energy difference between the CBM and the vacuum level, and since the introduced point defects move the CBM position further away from Fermi level, the energy difference (and the corresponding $\chi_{MoS_2}$ value) decreases. As can be seen from Figure S6 (c, d) (see Supplementary Materials), the introduction of point defects changes the Hartree potential profile, especially in their vicinity. Since the minimum value of the Hartree potential rises, the corresponding $\chi_{MoS_2}$ value, which is required to transfer an electron from the bottom of the conduction band to the vacuum level, is reduced. Thus, in the presence of point defects, the $\chi_{MoS_2}$, which is considered as an average over all possible sites of MoS$_2$ layer, including the defect sites, decreases. The magnitude of the effect depends not only on the type of defects but also their density.

As can be seen from Figure 7 (c), all the considered point defects, except for the AST defects, reduce the value of $\Delta V$ in the Au (111)/MoS$_2$ samples. The degree of reduction in $\Delta V$ depends on the type of point defects, which determines the value of interface dipole moment (see Figure 2 (d)). The smaller is the interface dipole, the smaller is the $\Delta V$. We found that the charges transfer from the Au substrate to the defect-free MoS$_2$, and therefore the resulting interface dipole moment is rather small, in agreement with [17,22]. The introduction of point defects further reduces the magnitude of interfacial dipole (see Figure S6 (b) in Supplementary Materials), and hence that of ΔV, ultimately leading to the larger values of SBH.

## The effect of defect density on SBH

Next, we investigate how the SBH depends on the defect density. To illustrate the effect of defect density on the electronic structure of MoS$_2$ monolayer placed on Au(111) surface, we plot the pDOS for Mo- and S-atoms of an MoS$_2$ monolayer with VT defects (see Figure 8 (a)) and AST defects (see Figure 8 (b)) at different defect densities. As can be seen in Figure 8 (a), the overall shape of pDOS for MoS$_2$ monolayer with VT defects varies insignificantly with the defect density. The height of the peak within the band gap (and to some extent its width) increases with an increase in the vacancy density. The distance between the Fermi level and the CBM, which is the measure of SBH, increases proportionally with the defect density.

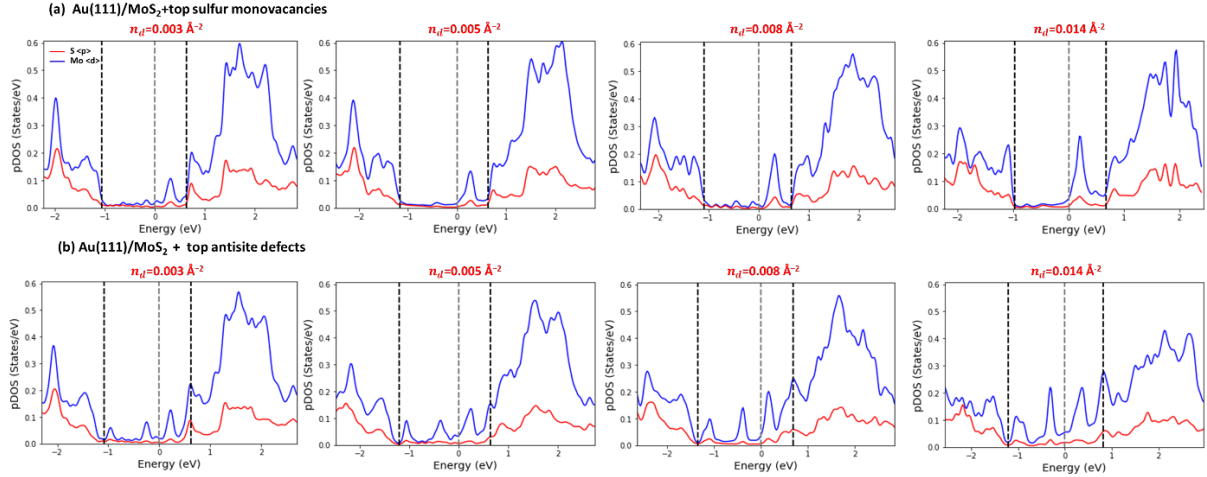

Figure 8: pDOS of Au/MoS$_2$ sample with the MoS$_2$ monolayer containing VT vacancies (a) and AST defects (b) at different defect densities. The pDOS calculated as an average over five d-orbitals of Mo-atoms indicated by blue, and over three p-orbitals of S-atoms indicated by red. The VBM, Fermi level and CBM, which were obtained with the projection method, are shown by the dashed lines.

The pDOS for Au(111) /MoS$_2$ sample with the MoS$_2$ monolayer containing AST defects are shown in Figure 8 (b) at different defect densities. Once again, one can see that at the different defect densities, the pDOS shape is similar: There are well-defined peaks located above and below the Fermi level in the band gap. It is evident that the height of the peaks grows, and their width broadens with the increase in defect density. The higher is the point defect density, the more distorted is the pDOS. Similar changes in the pDOS are found for the DV, VB, and AST defects (see Figures S1-S3 in Supplementary Materials).

The main effect of defect density on the SBH for different point defects is summarized in Figure 7 (d): The SBH monotonically increases with the defect density. An increase in the SBH is nearly linear for VT and AST defects, while it is strongly non-linear for ASB and DV defects, especially at high defect densities. In general, the higher the defect density is, the stronger its impact on the electronic structure of the Au(111)/MoS$_2$ contact is.

4. **Discussions**

Our study here shows that the results obtained by the two methods predict remarkably similar trends between the SBH and defect type, and between the SBH and defect density. On average, the difference in the SBH values is ~3%, while the maximal difference is only 7%. Interestingly, the difference in the SBH values obtained by these two methods in this study is smaller than previously reported [40]. This can be attributed to two reasons: first, we calculated the metal work function for the deformed Au(111) sample, and second, we used the contact MoS$_2$ monolayer instead of the free-standing one to calculate the corresponding electron affinity energy. Clearly, these two improvements make the method based on the SM rule more accurate.

We note that the experimentally measured values of SBH for MoS$_2$/Au junction fall in a broad range between 0.06 eV and 0.92 eV [1,8,38,39]. Our present study shows that the values of SBH can vary from 0.57 eV to 0.92 eV, depending on the type, density and location of point defects studied here. Hence, the present study can partially explain the large dispersion observed in experiments. In particular, the defect type and density play an important role. For example, the defect-free MoS$_2$/Au junction has the SBH of 0.57 eV and while that with ASB defects at a high density can lead to an SBH value of 0.92 eV.

It is understood that the MoS$_2$ samples used in experiments can be quite inhomogeneous, and the type, density, and location of defects in the samples can vary to a great extent, which can result in the large scattering of the SBH values. The previous study [1] has shown that the method used to fabricate the electrode to create the metal/MoS$_2$ junction can have a profound effect on the SBH. When a deposition method is used, the deposited ''high energy'' metal atoms can damage the lattice structure of MoS$_2$, which can lead to the substantial chemical disorders, namely formation of numerous S and Mo vacancies, and even metallic-like defects (metallic impurities) at the interface [3,15,42]. These chemical disorders can have a profound effect on the SBH. In particular, these metallic-like defects can lead to local Ohmic contacts, and thus can significantly reduce the overall SBH at the junction, which might explain the very low values of SBH observed in some of the experiments. In contrast, when atomically flat metal thin films are transferred onto MoS$_2$ monolayer (without direct chemical bonding) by using the damage-free electrode transfer technique [1], the observed interface is effectively free from chemical disorder, and this leads to much higher values of the measured SBH values. Hence, defect engineering, for example, by controlling the type, location, and defect density can play an effective role in modulating the SBH.

## 5. Conclusions

We performed first-principles calculations to investigate the effects of the type, location, and density of point defects in MoS$_2$ layer on the SBH of the Au (111)/MoS$_2$ junction. The values of SBH were calculated by applying two different methods: The first method is based on the projection of the electronic band structure and the second one is based on the SM rule. We found that these methods predict the same trend. With a couple of corrections, the two methods can lead to comparable values of SBH. Three types of point defects were studied: S monovacancy, S divacancy, and Mo$_S$ antisite defects. For S monovacancy and antisite defects, their presence in the top sublayer and bottom sublayer is differentiated. Overall, the SBH is sensitive to the type, density, and location of point defects in the MoS$_2$ layer. In general, the SBH for defective MoS$_2$ layer is universally higher than its defect-free counterpart, which will lead to a higher contact resistance and a lower electron injection efficiency. Among these defects, we found that the ASB and DV defects significantly increase the SBH, while the effect of VT, VB and AST defects is relatively weaker. Furthermore, the SBH monotonically increases with the defect density initially but gradually slows down. The effect of defect density for VT, VB and AST defects is smaller than that for ASB and DV defects.

The present work suggests that the reported dispersion of the experimentally measured SBH values for Au/MoS$_2$ junction can be at least partially accounted by the existence of point defects in MoS$_2$ monolayer. The present study also suggests that the SBH can be modulated via defect engineering of MoS$_2$ layer, for example, by controlling the type, location, or density of defects. Hence, our findings can serve as a guide for the control and optimization of the SBH in Au/MoS$_2$ heterojunctions.

## 6. Acknowledgements

This work was supported by the National Research Foundation, Singapore under Award No. NRF-CRP24-2020-0002. Zhang Y.W. acknowledges the support from Singapore A*STAR SERC CRF Award. The use of computing resources at the A*STAR Computational Centre and National Supercomputer Centre, Singapore is gratefully acknowledged.

## 6. Citations

## Supplementary Materials

# The effects of point defect type, location, and density on the Schottky barrier height of Au/MoS2 heterojunction: A first-principles study

Viacheslav Sorkin[1,§], Hangbo Zhou[1], Zhi Gen Yu[1], Kah-Wee Ang[2,3,**], Yong-Wei Zhang[1,††]

[§] Email: sorkinv@ihpc.a-star.edu.sg
[**] Email: kahwee.ang@nus.edu.sg
[††] Email: zhangyw@ihpc.a-star.edu.sg

[1]Institute of High-Performance Computing, A*STAR, 1 Fusionopolis Way, Singapore 138632

[2]Department of Electrical and Computer Engineering, National University of Singapore, 4 Engineering Drive 3, Singapore, 117583

[3]Institute of Materials Research and Engineering, A*STAR, 2 Fusionopolis Way, Singapore, 138634

## Data from DFT calculations

In this section, we report the data obtained from our DFT calculations with the method based on the projection of electronic band structure and the modified SM rule. The calculated values of Schottky barrier height (SBH), electron affinity energy (EAE), and potential step, ΔV, for Au(111)/MoS$_2$ heterojunction with a defect-free monolayer as well as MoS$_2$ layer with top and bottom single S-vacancies, double S-vacancies, top and bottom anti-site Mo$_S$ defects are reported in Tables S1-S4. The data are obtained by using PBE exchange-correlation functional with and without van der Waals DFT-D2 corrections. The obtained data are presented in Table S1 for the 6x6x4 Au(111)/MoS$_2$ sample, those in Table S2 for the 6x5x4 Au(111)/MoS$_2$ sample, those in Table S3 for the 4x4x4 Au(111)/MoS$_2$ sample, and those in Table S4 for the 3x3x6 Au(111)/MoS$_2$ sample.

Table S2: The calculated values of Schottky barrier height (SBH), electron affinity energy (EAE), and potential step, ΔV, for Au(111)/MoS$_2$ contact with a defect-free monolayer (PF) and one containing top (VT) and bottom (VB) single S-vacancies, double S-vacancies (DV), as well as top (ST) and bottom (SB) anti-site Mo$_S$ defects are reported. The SBH are calculated by the method based on electronic band structure projection (SBH-PJ) and modified Schottky-Mott rule (SBH-HP). The data for the 6x6x4 Au(111)/MoS$_2$ samples are calculated by using PBE exchange-correlation functional with and without van der Waals DFT-D2 corrections.

| Defect type | PBE | | | | | PBE + van der Waals corrections | | | | |
|---|---|---|---|---|---|---|---|---|---|---|
| | SBH-PJ (eV) | SBH-HP (eV) | Difference (%) | EAE (eV) | ΔV (eV) | SBH-PJ (eV) | SBH-HP (eV) | Difference (%) | EAE (eV) | ΔV (eV) |
| PF | 0.67 | 0.69 | 3 | 4.20 | 0.21 | 0.57 | 0.60 | 5 | 4.35 | 0.32 |
| VT | 0.73 | 0.76 | 4 | 4.16 | 0.18 | 0.64 | 0.66 | 3 | 4.32 | 0.29 |

| | | | | | | | | | |
|---|---|---|---|---|---|---|---|---|---|
| VB | 0.74 | 0.75 | 3 | 4.16 | 0.19 | 0.64 | 0.66 | 3 | 4.32 | 0.29 |
| DV | 0.77 | 0.78 | 2 | 4.14 | 0.18 | 0.68 | 0.71 | 5 | 4.26 | 0.30 |
| AST | 0.73 | 0.75 | 2 | 4.06 | 0.30 | 0.63 | 0.67 | 6 | 4.20 | 0.40 |
| ASB | 0.88 | 0.89 | 2 | 4.04 | 0.17 | 0.78 | 0.78 | 0 | 4.20 | 0.29 |

Table S3: The calculated values of Schottky barrier height (SBH), electron affinity energy (EAE), and potential step, ΔV, for Au(111)/MoS$_2$ contact with a defect-free monolayer (PF) and one containing top (VT) and bottom (VB) single S-vacancies, double S-vacancies (DV), as well as top (ST) and bottom (SB) anti-site defects are reported. The SBH are calculated by the method based on electronic band structure projection (SBH-PJ) and modified Schottky-Mott rule (SBH-HP). The data for the 5x5x4 Au(111)/MoS$_2$ samples are calculated by using PBE exchange-correlation functional with and without van der Waals DFT-D2 corrections.

| Defect type | PBE | | | | | PBE + van der Waals corrections | | | | |
|---|---|---|---|---|---|---|---|---|---|---|
| | SBH-PJ (eV) | SBH-HP (eV) | Difference (%) | EAE (eV) | ΔV (eV) | SBH-PJ (eV) | SBH-HP (eV) | Difference (%) | EAE (eV) | ΔV (eV) |
| PF | 0.67 | 0.69 | 3 | 4.20 | 0.21 | 0.57 | 0.61 | 7 | 4.36 | 0.30 |
| VT | 0.77 | 0.79 | 3 | 4.14 | 0.17 | 0.67 | 0.70 | 4 | 4.31 | 0.26 |
| VB | 0.78 | 0.80 | 3 | 4.14 | 0.16 | 0.67 | 0.69 | 3 | 4.31 | 0.27 |
| DV | 0.82 | 0.84 | 3 | 4.10 | 0.16 | 0.71 | 0.75 | 6 | 4.25 | 0.27 |
| AST | 0.75 | 0.78 | 5 | 4.02 | 0.30 | 0.65 | 0.67 | 3 | 4.20 | 0.40 |
| ASB | 0.92 | 0.94 | 3 | 4.00 | 0.16 | 0.82 | 0.84 | 2 | 4.19 | 0.24 |

Table S4: The calculated values of Schottky barrier height (SBH), electron affinity energy (EAE), and potential step, ΔV, for Au(111)/MoS$_2$ contact with a defect-free monolayer (PF) and one containing top (VT) and bottom (VB) single S-vacancies, double S-vacancies (DV), as well as top (ST) and bottom (SB) anti-site defects are reported. The SBH are calculated by the method based on electronic band structure projection (SBH-PJ) and modified Schottky-Mott rule (SBH-HP). The data for the 4x4x4 Au(111)/MoS$_2$ samples are calculated by using PBE exchange-correlation functional with and without van der Waals DFT-D2 corrections.

| Defect type | PBE | | | | | PBE + van der Waals corrections | | | | |
|---|---|---|---|---|---|---|---|---|---|---|
| | SBH-PJ (eV) | SBH-HP (eV) | Difference (%) | EAE (eV) | ΔV (eV) | SBH-PJ (eV) | SBH-HP (eV) | Difference (%) | EAE (eV) | ΔV (eV) |
| PF | 0.67 | 0.69 | 3 | 4.20 | 0.21 | 0.57 | 0.59 | 4 | 4.36 | 0.31 |
| VT | 0.81 | 0.83 | 2 | 4.11 | 0.16 | 0.71 | 0.72 | 1 | 4.30 | 0.24 |
| VB | 0.82 | 0.83 | 2 | 4.11 | 0.16 | 0.71 | 0.71 | 0 | 4.30 | 0.24 |
| DV | 0.87 | 0.88 | 1 | 4.06 | 0.16 | 0.76 | 0.81 | 7 | 4.23 | 0.26 |
| AST | 0.77 | 0.81 | 6 | 3.97 | 0.33 | 0.68 | 0.71 | 5 | 4.18 | 0.40 |
| ASB | 0.97 | 1.00 | 3 | 3.96 | 0.15 | 0.87 | 0.93 | 6 | 4.18 | 0.21 |

Table S5 The calculated values of Schottky barrier height (SBH), electron affinity energy (EAE), and potential step, ΔV, for Au(111)/MoS$_2$ contact with a defect-free monolayer (PF) and one containing top (VT) and bottom (VB) single S-vacancies, double S-vacancies (DV), as well as top (ST) and bottom (SB) anti-site defects are reported. The SBH are calculated by the method based on electronic band structure projection (SBH-PJ) and modified Schottky-Mott rule (SBH-HP). The data for the 3x3x6 Au(111)/MoS$_2$ samples are calculated by using PBE exchange-correlation functional with and without van der Waals DFT-D2 corrections.

| Defect type | PBE | | | | | PBE + van der Waals corrections | | | | |
|---|---|---|---|---|---|---|---|---|---|---|
| | SBH-PJ (eV) | SBH-HP (eV) | Difference (%) | EAE (eV) | ΔV (eV) | SBH-PJ (eV) | SBH-HP (eV) | Difference (%) | EAE (eV) | ΔV (eV) |
| PF | 0.67 | 0.72 | 7 | 4.18 | 0.20 | 0.57 | 0.61 | 7 | 4.34 | 0.32 |

| | | | | | | | | | |
|---|---|---|---|---|---|---|---|---|---|
| VT | 0.84 | 0.85 | 1 | 4.10 | 0.15 | 0.75 | 0.78 | 4 | 4.28 | 0.21 |
| VB | 0.85 | 0.86 | 1 | 4.09 | 0.15 | 0.75 | 0.78 | 4 | 4.28 | 0.21 |
| DV | 0.90 | 0.92 | 3 | 4.04 | 0.14 | 0.80 | 0.83 | 4 | 4.20 | 0.24 |
| AST | 0.82 | 0.85 | 4 | 3.92 | 0.33 | 0.71 | 0.73 | 2 | 4.16 | 0.38 |
| ASB | 1.00 | 1.01 | 1 | 3.92 | 0.14 | 0.91 | 0.93 | 3 | 4.16 | 0.18 |

## Effect of defect density on SBH

We plot the pDOS for Mo- and S-atoms of an MoS$_2$ monolayer with bottom single S-vacancies, double S-vacancies, and bottom antisite Mo$_S$ defects at various defect densities (per unit area of MoS$_2$ ) in Figure S1, Figure S2 and Figure S3, respectively. As can be seen in Figure S1, the overall shape of pDOS for the MoS2 monolayer with bottom S-monovacancies varies with the defect density: the height of peak within the band gap increases proportionally with the defect density. The similar changes in the pDOS Au(111)/MoS$_2$ monolayer with S-divacancies with defect density are shown in Figure S2: the height of peaks in the band gap increases. An additional peak close to the bottom of conduction band appears at the highest vacancy density.

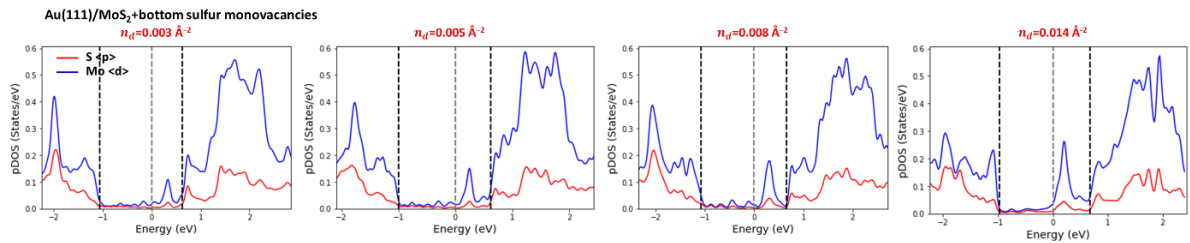

Figure S9: PDOS of Au(111) /MoS2 sample with a monolayer containing bottom single sulfur vacancies at different defect densities. The pDOS calculated as an average over five d-orbitals of Mo-atoms indicated by blue, and over three p-orbitals of S-atoms indicated by red. The VBM, Fermi level and CBM, obtained with the PJ method, are shown by dashed lines. The PDOS are for DFT calculations with DT2 van der Waals corrections.

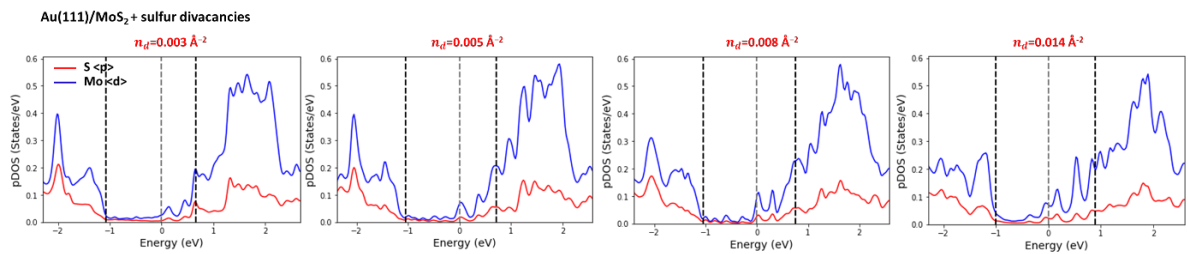

Figure S10: PDOS of Au(111)/MoS2 sample with a monolayer containing bottom double sulfur vacancies at different defect densities. The pDOS calculated as an average over five d-orbitals of Mo-atoms indicated by blue, and over three p-orbitals of S-atoms indicated by red. The VBM, Fermi level and CBM, obtained with the PJ method, are shown by dashed lines. The PDOS are for DFT calculations with DT2 van der Waals corrections.

The pDOS for of Mo- and S- atoms in the Au(111)/MoS$_2$ sample with a monolayer containing bottom anti-site defects at different defect densities are shown in Figure S3. The interaction of the anti-site defects with the underlying gold substrate is stronger as compared with the other studied defects. As a result, many states appear in the band gap, forming the broad continuous-like spectrum of states as shown in Figure S3. The density of the introduced states increases in direct proportion with defect concentration.

However, the CBM position is clearly visible in the pDOS, and it can be used to demonstrate the increase in SBH with defect concentration.

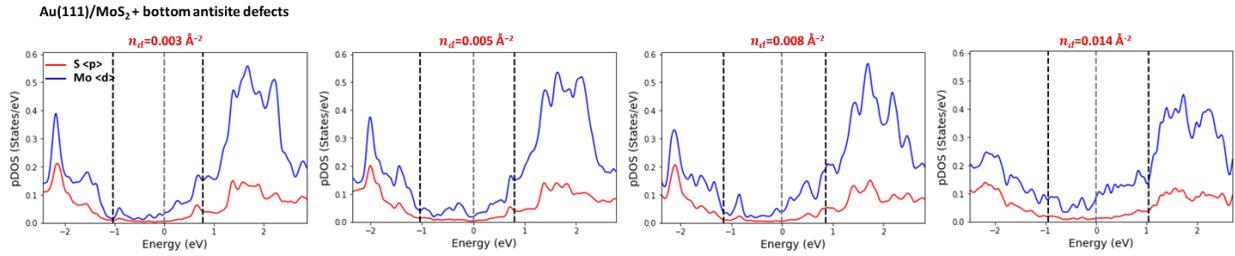

**Figure S11: pDOS of Au(111) /MoS$_2$ sample with a monolayer containing bottom anti-site defects at different defect densities. The PDOS calculated as an average over five d-orbitals of Mo-atoms indicated by blue, and over three p-orbitals of S-atoms indicated by red. The VBM, Fermi level and CBM, obtained with the PJ method, are shown by dashed lines.**

The effect of defect density per unit area on the SBH obtained by DFT calculations with PBE XC-functional by using the method based on projection of electronic band structure is reported in Figure S4 (a). The SBH values for Au/MoS$_2$ sample with a MoS$_2$ monolayer containing single top (blue circles) and bottom (magenta circles) sulfur vacancies, double sulfur vacancies (green triangles), as well as top (black squares) and bottom (red squares) anti-site Mo$_S$ defects are shown in Figure S4 (b). The SBH value of the Au(111)/MoS$_2$ heterojunction with a defect-free MoS$_2$ is given for comparison.

## The effect of different point defects on the SBH

The effect of point defects on SBH is illustrated in Figure S4 (a). The SBH value for Au(111)/MoS$_2$ heterojunction with a defect-free MoS$_2$ monolayer and a monolayer containing single top and bottom sulfur monovacancies, sulfur divacancies, as well as top and bottom anti-site Mo$_S$ defects are shown in Figure S4 (b). The DFT calculations with PBE XC-functional are reported obtained by the PJ method based on projection of electronic band structure.

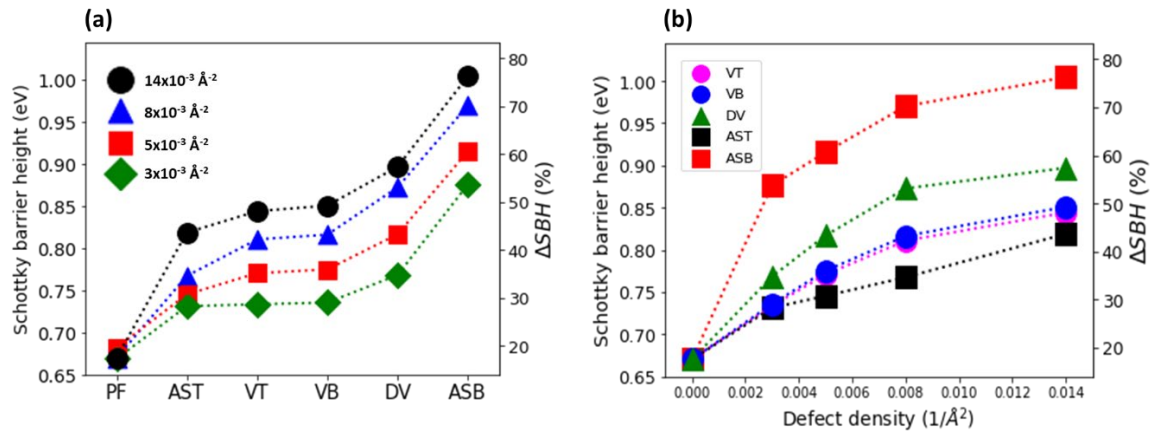

Figure S12: (a) The effect of concentration of point defects on the SBH. The SBH value for Au(111)/MoS$_2$ contact with a defect-free monolayer (PF) and a monolayer containing single top (VT) and bottom (VB) sulfur vacancies, double sulfur vacancies (DV), as well as top (AST) and bottom (ASB) anti-site defects. The data is from 3x3 samples with gold 6 layers (black circles), and 5x5 (red squares), 4x4 (blue triangles), and 6x6 (green diamonds) Au/MoS$_2$ samples with gold 4 layers. (b) The effect of defect concentration on the SBH for the Au(111)/MoS2 samples with a MoS$_2$ monolayer containing single top (blue circles) and bottom (clue circles) sulfur vacancies, double sulfur vacancies (green triangles), as well as top (black squares) and bottom (red squares) anti-site defects. The SBH value of the defect-free sample is given for comparison. The reported SBH values were obtained by DFT calculations with PBE XC-functional using the method based on projection of electronic band structure.

Table S6: Increase in the value of SBH in defective MoS$_2$ monolayer. The reported data for DFT calculations with PBE and PBE with van der Waals corrections. The electronic band structure projection method is used.

| Defect type | Δ SBH (%) {PBE} | | | | Δ SBH (%) {PBE + van der Waals corrections} | | | |
|---|---|---|---|---|---|---|---|---|
| | $n_d$=0.003 (1/Å$^2$) | $n_d$=0.005 (1/Å$^2$) | $n_d$=0.008 (1/Å$^2$) | $n_d$=0.014 (1/Å$^2$) | $n_d$=0.003 (1/Å$^2$) | $n_d$=0.005 (1/Å$^2$) | $n_d$=0.008 (1/Å$^2$) | $n_d$=0.014 (1/Å$^2$) |
| VT | 9% | 13% | 21% | 26% | 13% | 18% | 25% | 32% |
| VB | 10% | 14% | 22% | 27% | 13% | 17% | 25% | 32% |
| DV | 15% | 20% | 30% | 34% | 18% | 24% | 33% | 40% |
| AST | 9% | 10% | 15% | 22% | 11% | 14% | 19% | 25% |
| ASB | 31% | 35% | 45% | 50% | 37% | 44% | 53% | 59% |

## Comparison of the two calculation methods and the effect of van der Waals corrections

Figure S5 (a-d) compare the SBH values calculated with the method based on projection of electronic band structure of MoS$_2$ monolayer on band structure of Au(111)/MoS2 contact (blue circles) and with the method based on Hartree potential and the modified Schottky-Mott rule (red squares). In Table S5, we compare the SBH values calculated with the PBE exchange-correlation (XC) potential and PBE-XC with van der Waals DFT-D2 corrections obtained by the method based on projection of electronic band structure.

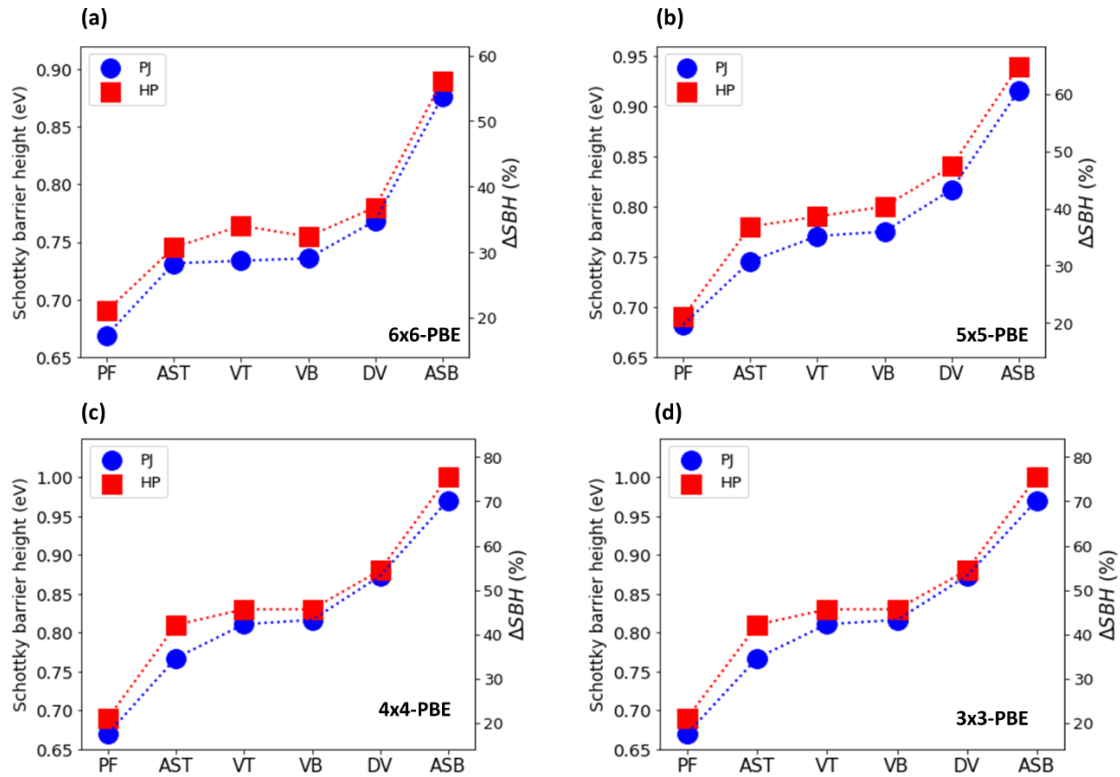

Figure S13: Comparison of the SBH values calculated with the method based on projection (PJ) of electronic band structure of MoS2 monolayer on band structure of Au/MoS2 contact (blue circles) and with the method based on Hartree potential (HP) and the modified Schottky-Mott rule (red squares) for (a) 6x6x4, (b) 5x5x4, (c) 4x4x4 and (d) 3x3x6 Au(111)/MoS$_2$ sample with PBE XC-functional.

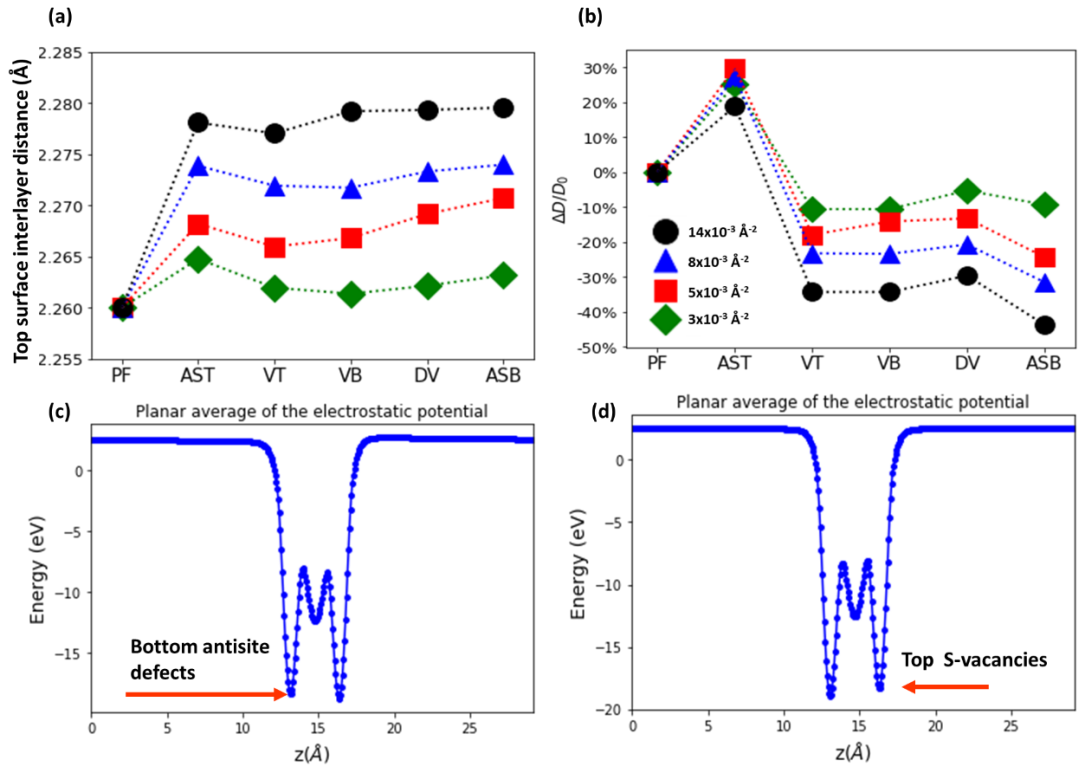

Figure S14: (a) Interlayer distance between the two top adjacent layers of Au(111) surface in contact with defect-free (PF) and defective MoS$_2$ monolayer. Defective monolayer contains single top (VT) and bottom (VB) sulfur vacancies, double sulfur vacancies (DV), as well as top (AST) and bottom (ASB) antisite defects. The defect densities are indicated in the inset (b). Variation of the interfacial dipole moment of the defective MoS$_2$ layer with respect to the value of defect-free monolayer: $\frac{\Delta D}{D_0} = 100 * \left(\frac{D-D_0}{D_0}\right)$ (c, d) Planar average of Hartree potential for Au(111)/MoS$_2$ sample with the MoS$_2$ monolayer containing a bottom antisite defect (c) and a single top vacancy (d). The Z-axis is normal to the Au(111)/MoS2 interface; the plane average is calculated over [XY] planes along the sample. The plots are for the Au(111)/MoS2 6x6x4 samples.

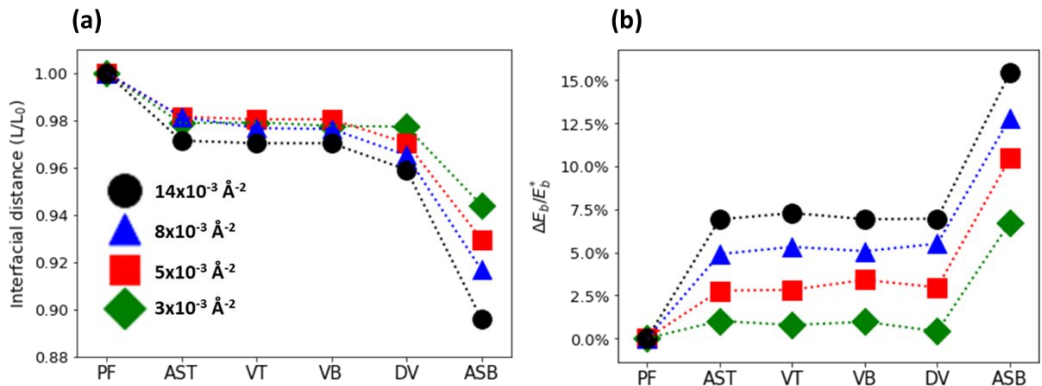

Figure S15: (a) Interfacial distance for the Au(111)/MoS$_2$ sample with defect-free (PF) and defective MoS$_2$ monolayer. Defective monolayer contains single top (VT) and bottom (VB) sulfur vacancies, double sulfur vacancies (DV), as well as top (AST) and bottom (ASB) antisite defects. The defect densities are indicated in the inset. (b) Variation of the binding energy, $E_b$, of the defective MoS$_2$ layer with respect to the value of defect-free monolayer, $E_b^*$: $\frac{\Delta E_b}{E_b^*} = 100 * \left(\frac{E_b - E_b^8}{E_b^*}\right)$.